\documentstyle[aaspp4,epsfig]{article}
 \def\simlt{\lower.5ex\hbox{$\; \buildrel < \over \sim \;$}}
  \def\simgt{\lower.5ex\hbox{$\; \buildrel > \over \sim \;$}}
\def\mag{\mbox{ mag}}
\def\kms{\mbox{ km s$^{-1}$}}
\def\mpc{\mbox{ Mpc}}
\def\kpc{\mbox{ kpc}}

\def\msun{\mbox{ M}_\odot}

\begin{document}
\title{Probing the Structure of Dark Matter in Galaxy Halos and Clusters
using Supernovae}

\author{R. Benton Metcalf}
\affil{\it Institute of Astronomy, University of Cambridge \\ Madingley Road
Cambridge CB3 0HA, UK} 

\begin{abstract}
A new method for measuring gravitational lensing with high redshift type
Ia supernovae is investigated.  
The method utilizes correlations 
between foreground galaxies and supernova brightnesses to substantially
reduce possible systematic errors and increase the signal to noise ratio.  
It is shown that this lensing signal can be related to the mass, size and
substructure of galaxy halos and galaxy clusters if dark matter consists
of microscopic particles.  This technique may be
particularly useful for measuring the size of dark matter halos, a
measurement to which the lensing of galaxies is not well suited, and for
measuring the level of substructure in galaxy halos, a problematic
prediction of the cold dark matter model.  The contributions to the
signal from galaxy halos and galaxy clusters are modeled and
contributions to the noise from fluctuations in the galaxy number
counts, galaxy redshift error, dispersion in SN luminosities and sample
variance are estimated.  The intrinsic distribution of
supernova luminosities and its redshift evolution are removed as major
sources of uncertainty.  The method is found to be complimentary to
galaxy--galaxy lensing.  The required observations of $\simgt 100$
supernovae have already been proposed for the purposes of cosmological
parameter estimation.
\end{abstract}

\keywords{Cosmology, Dark Matter, Gravitational Lensing, Supernovae}
\section{Introduction}

Outside of the visible extent of galaxies not a great deal is known
about the distribution of dark matter (DM) on scales smaller than galaxy
clusters.  Galaxy--galaxy lensing
\markcite{astro-ph/9912119,1998ApJ...503..531H, 1996ApJ...466..623B,
1996MNRAS.282.1159G, 1984ApJ...281L..59T}(Fischer {et~al.} 1999; {Hudson} {et~al.} 1998; {Brainerd}, {Blandford}, \&  {Smail} 1996; {Griffiths} {et~al.} 1996; {Tyson} {et~al.} 1984) and studies of
the satellite galaxies orbiting the Milky Way and other galaxies
\markcite{1999MNRAS.310..645W,1996ApJ...457..228K,1994ApJ...435..599Z}({Wilkinson} \& {Evans} 1999; {Kochanek} 1996; {Zaritsky} \& {White} 1994) 
 have put some constraints on the mass of galaxy
halos, but their size scale and the degree to which they are smooth mass
distributions or collections of subclumps are not well determined.  Even
the concept of galaxy halos, with a single galaxy within each of them and well
defined sizes, may not be correct.  Nor is it clear how the observable
properties of galaxies relate to the DM around them.  It would be
preferably if these properties could be derived directly from cosmological
initial conditions and the properties of DM, but this has proven
to be an extremely difficult problem.  Simulations of galaxy formation give
incomplete answers which are hard to compare with observations.
For example pure cold dark matter (CDM) simulations contain a large
amount of small scale structure which is not mirrored in the observed
distribution of light \markcite{1999ApJ...524L..19M}({Moore} {et~al.} 1999a).  The process of
galaxy formation is  highly dependent on cooling and feedback from stars
which must be modeled ad hoc.  As a result complete galaxies have never
been created in a hydrodynamic simulation.  In addition, changing the
properties of DM so that it is self--interacting or a mixture of hot and
cold DM, for example, can have a strong influences on small scale structure.  
The theory of galaxy formation, and DM, would clearly benefit from further
observational constraints on these scales.  Gravitational lensing is
presently the most promising method available for probing the
distribution of dark matter and its relationship with light.

The discovery of a tight correlation between the light curve
shape and the luminosities of type Ia supernovae (SNe) has resulted in their
great utility as cosmological distance indicators \markcite{hamuy96,RPK96}({Hamuy} {et~al.} 1996; {Riess}, {Press}, \& {Kirshner} 1996)
and as a possible probe of gravitational magnification.
The measured standard deviation of corrected peak luminosities in local
SNe is now $\simlt 0.12 \mag$.
Two groups have used these SNe to constrain the
distance--redshift relation and through this cosmological 
parameters \markcite{1997ApJ...483..565P,1998AJ....116.1009R}({Perlmutter} {et~al.} 1997; {Riess} {et~al.} 1998).  The
successes of these studies have inspired plans for more aggressive,
larger searches for high redshift SNe in the near future.  The volume and
quality of data is likely to increase dramatically.
These data will have implications well beyond constraints on $\Omega_m$ and
$\Omega_\Lambda$ -- the average mass density and vacuum energy density.  
Other uses include measuring the universe's equation
of state and star formation history.  Here I discuss the
gravitational lensing of high redshift SNe and its implications for dark
matter, its distribution and its correlation with light.

\markcite{1999ApJ...519L...1M}{Metcalf} \& {Silk} (1999) have already discussed using the lensing of SNe to
determining the composition of DM.  It was found that DM made predominantly
of macroscopic compact objects with masses $\simgt 10^{-3}\msun$ (ex. MACHOs, primordial black holes) can be
differentiated from DM made of microscopic particles (ex. WIMPS, axions)
through studying the magnification probability distribution of $\sim
100$ SNe at $z\sim 1$.  In the case of macroscopic DM, the most likely
brightness of a supernova (SN) is well below the mean value, and near the empty
beam solution or Dyer--Roeder distance \markcite{1974ApJ...189..167D}({Dyer} \& {Roeder} 1974).  
\markcite{1999A&A...351L..10S}{Seljak} \& {Holz} (1999) expanded on this work by considering a
mixture of macroscopic and microscopic DM.  Unknown to the author,
\markcite{1991ApJ...374...83R}{Rauch} (1991) had previously suggest this use for type Ia SNe, but had not investigated the shift in the mode of the distribution to
values below the standard Friedmann-Robertson-Walker luminosity
distance.  Instead he concentrated on the detection of rare high
magnification events and concluded that many more SNe would be require
to detect macroscopic compact objects.  In this paper only microscopic
DM is considered.  In this case the DM can be treated as a transparent,
massive fluid clumped into structures that are much larger than the
angular size of the source.

A few authors have considered how the lensing of type Ia SNe could be used to 
probe DM structure.  \markcite{kolatt98}{Kolatt} \& {Bartelmann} (1998) pointed out that the mass sheet
degeneracy of galaxy cluster shear maps could be broken with SNe. 
\markcite{Madau99}{Porciani} \& {Madau} (2000) calculated the probability of strong lensing events
in a CDM universe.  They used the
Press-Schechter prediction for the mass function of halos and find the
optical depth for a certain magnification threshold.  High magnification
events are rare ($\sim 2\times 10^{-3}$ chance for magnification above
$1.3$ at $z_s=1$) although they could be 
specifically searched for by monitoring high--mass galaxy clusters.  The
approach discussed in this paper incorporates both strongly and weakly
lensed SNe to take better advantage of the data and broaden the
sensitivity to mass outside halo cores hence probing the size scale of 
the extended DM distribution.
The author has previously pointed out that lensing by microscopic DM
could be detected by an increase in the variance of SN brightnesses at
high redshift \markcite{1999MNRAS.305..746M}({Metcalf} 1999).  In addition this increase
in dispersion is important for determining the noise in cosmological
parameter estimates when using SNe at $z\sim 1$.  The  
feasibility of measuring lensing via this method is potentially hampered by
uncertainties in the intrinsic distribution of SN luminosities and its
possible redshift dependence.  In this paper, I seek to eliminate these
systematic errors by correlating SN brightnesses with foreground
galaxies.   In this 
way a null result is expected in the absence of lensing.  This approach
also makes interpretation of the result more straightforward and results
in significantly higher signal to noise.  This work is an extension and
improvement an earlier parametric approach summarized in
\markcite{1998elss.confE..25M}Metcalf (1998).

In section~\ref{sec:Correlating-SNe-with} the first order correlation
statistic is defined.  In section~\ref{sec:Model-Sign-Noise} models are
constructed for estimating and interpreting the contribution of galaxy
halos~(\ref{sec:Galaxy-Halos}) and extragalactic
clustering~(\ref{cluster_component}) to the lensing signal.  Sources of
noise are considered in section~\ref{sec:noise}.  Some higher order
correlation statistics are discussed in
section~\ref{sec:High-Order-Corr}.  The lensing of SNe is compared to
the lensing of galaxies in section~\ref{sec:Comp-with-lens}.  The final
section contains a summary and discussion of future prospects.

\section{Correlating SNe with Foreground Light}
\label{sec:Correlating-SNe-with}

As stated above I seek to reduce as much as possible the
influence of SN redshift evolution by considering only correlations in
foreground light and SN brightness.  Using statistics based on these
correlations also increases the signal to noise and allows for simple
interpretations of the results in terms of mass--light correlations.
Throughout this paper units are used in which the speed of light, $c=1$.

To start, consider the weighted flux from foreground galaxies
\begin{eqnarray}
{\mathcal F} = \sum_i w(z_i,z_s,y_i) f_i
\end{eqnarray}
where $y_i$ is the proper distance separating the galaxy $i$ from the
line--of--sight to the SN, $f_i$ is the flux received from that galaxy
and $z_s$ is the redshift of the source SN.  An attempt should be made
to correct the $f_i$'s for extinction and inclination effects as well as
k-corrections so that $f_i$ accurately reflects the intrinsic
luminosity of the galaxy.
I will choose the weight function to be factorable and normalized as
follows
\begin{eqnarray}
\begin{array}{c}
w(z,z_s,y) = w_1(z,z_s)w_2(y) \\
\int_0^{z_s} dz~ w_1(z,z_s) = 1 ~~,~~  2\pi\int_0^{\infty} dy~y~w_2(y) = 1.
\end{array}
\end{eqnarray}
This gives ${\mathcal F}$ the units of flux per area.  Besides these
constraints the weighting function is arbitrary so that it can be
adjusted to maximize the signal to noise.  Later I will make a
particular choice of weighting function for numerical calculations.

For simplicity I will consider a flux limited sample of foreground
galaxies.  In this case the average of ${\mathcal F}$ is
\begin{eqnarray}
\overline{\mathcal F} = \int_0^{z_s} dz~w_1(z,z_s) \frac{\overline{f}(z) \eta(z)}{(1+z)}\frac{d\chi}{dz}(z) = 
\frac{1}{4\pi H_o} \int_0^{z_s} dz~w_1(z,z_s) \frac{(1+z)}{E(z)
d_L(z)^2} \int_{0}^\infty dL~L \phi(L,z) \Phi\left(L,z\right)
\end{eqnarray}
\begin{eqnarray}
E(z)\equiv \left[ \Omega_m(1+z)^3 + (1-\Omega_m-\Omega_\Lambda)(1+z)^2
+\Omega_\Lambda \right]^{1/2} ~~~~,~~~~~~
\chi(z)=\frac{1}{H_o}\int^z_0 \frac{dz'}{E(z')}
\end{eqnarray}
where $\eta(z)$ is the number density (number per proper volume) of
observed galaxies, $d_L(z)$ is the
luminosity distance, $\phi(L,z)$ is the selection function of the survey
and $\Phi(L,z)$ is the luminosity function in comoving volume.  The
comoving radial distance is $\chi(z)$.  The
selection function, $\phi(L,z)$, will depend on details of the
particular survey.  In the numerical calculations shown later a simple
flux limited survey will be assumed where $\phi(L,z)$ zero for
$L<L_c(z)$ and one for larger $L$.  In terms of the flux limit
$L_c(z)=4\pi d_L(z)^2 f_c$.
The luminosity function of detectable galaxies, $\phi(L,z)
\Phi\left(L,z\right)$, can be found empirically and independently of
the SNe.  

A SN is an unresolved near--point source so the only observable effects of
gravitational lensing are image magnification and the multiplication of images.
If DM is microscopic as is assumed here the probability of a SN at
$z\simeq 1$ having multiple, distinguishable images is very small
($< 10^{-4}$).  Except for rare cases, SNe will be only weakly magnified
and demagnified by matter very close to the line--of--sight and no
multiple images will occur.  To second order the magnification, $\mu = 1+2\kappa+3\kappa^2+\gamma^2$.  The convergence, $\kappa$, is defined by
its relation to the magnification matrix: $\mbox{tr}A=2(1-\kappa)$ and
the shear is $\gamma^2=|A-(1-\kappa) I|$.  So in the weak lensing limit,
$\kappa,\gamma \ll 1$, the magnification is independent of the shear.
The convergence, and thus $\mu$ in this case, depends only on the local
mass density (Ricci focusing 
only) and can be calculated perturbatively \markcite{Kais92,blan91}({Kaiser} 1992; {Blandford} {et~al.} 1991)
\begin{eqnarray}
\mu-1=\delta\mu \simeq \int^{z_s}_0 dz~ W_l(z,z_s)
\delta(z)
~~~~,~~~~
W_l(z,z_s) \simeq \frac{3\Omega_m H_o (1+z)}{E(z)}\frac{g\left[\chi(z)\right]g\left[\chi(z_s)-\chi(z)\right]}{g\left[\chi(z_s)\right]}
\end{eqnarray}
where the density contrast along the line--of--sight is
$\delta(z)=(\rho-\overline{\rho})/\overline{\rho}$.  The comoving
angular size distance is $g(\chi)=\left\{
R_{curv}\sin(\chi/R_{curv}),\chi,R_{curv}\sinh(\chi/R_{curv})
\right\}$ in the closed, flat and open cases.  The curvature scale is
$R_{curv}^{-1}=H_o\sqrt{|1-\Omega_m-\Omega_\Lambda |}$.
This can be easily modified in everything that follows to account for
strong lensing as well as weak.  In this notation the luminosity
distance is $d_L(z)=(1+z)g[\chi(z)]$.

\subsection{Correlations}
\label{sec:Correlations}

Consider the cross-correlation of $\mathcal F$ with the
brightness of the SN, $b$.  At a given redshift the magnification is
defined with respect to the mean brightness at that redshift.  
Ideally the average brightness of the standard candle,
$\overline{b}(z)$, would be calculated from the SNe themselves.
This would make the measurement of lensing independent
of cosmology and any redshift evolution of the SNe.  However this may
not prove feasible in the short term if there are not enough SNe within
a small redshift range to accurately determine the average.   In this
case the cosmology can be fixed so that the redshift dependence of the
mean brightness is predicted.  Determining the cosmological parameters
can be effectively done with lower redshift SNe which are largely
unaffected by lensing and with other observations.  
The average $\overline{\mathcal F}(z)$ can be measured without regard to
the presence of background SNe.

I will consider the correlation function
\begin{eqnarray}
{\mathcal S}^2_{{\mathcal F} \mu} \equiv \frac{1}{N_{sn}}\sum_{j} \frac{  \Delta{\mathcal
F}_j \Delta b_j}{\overline{\mathcal F}_j~\overline{b}_j }
\end{eqnarray}
where $\Delta b_i=b_i-\overline{b}(z_i)$.  The sum is over the sample of
SNe.  If the SNe are used to estimate the average $\overline{b}(z_i)$ one must
take into account the correlations in the finite sample.
The averages of these statistics can be expressed in terms of the
magnification of the SNe in a redshift range
\begin{eqnarray}
\left\langle {\mathcal S}^2_{{\mathcal F} \mu} \right\rangle = \left\langle
\delta{\mathcal F} \delta\mu \right\rangle.
\end{eqnarray}
where $\delta{\mathcal F}=\left({\mathcal F}-\overline{\mathcal F}\right)/\overline{\mathcal F}$.

To interpret and predict measurements of this statistic, its mean
value must be related to the distribution of DM and galaxies.  
For these purposes I define a position--dependent luminosity function by
writing the probability of a galaxy being within a comoving volume
$\delta V$ and in a range of luminosity $\delta L$ as
\begin{eqnarray}\label{galaxy_prob}
\delta V \delta L ~\Phi(\vec{x},L,z) = \delta V \delta L
~\Phi(L,z)\left[1+\delta\Phi\left( \vec{x},L;\delta_{\vec{y}} \right) \right].
\end{eqnarray}
The probability of a galaxy being at $\vec{x}$ is presumably
dependent on the density field, $\delta_{\vec{y}}$, at the point
$\vec{x}$ and in surrounding regions.  This luminosity function is a
{\em function} of position and luminosity, but a {\em functional} of the
density field.  This functional property expresses the probability that a
galaxy's existence and properties depend on more than the
smoothed density at its position.  Nonlocal properties of the
surrounding density field are probably important.

In relating these statistics to correlations of galaxies and mass
density I will employ an extended Limber's approximation which requires
that all such correlations are limited to scales much smaller than those
over which quantities like redshift, the lensing window and the angular size
distance change appreciably.  With this approximation and the luminosity
function above, the correlations between magnification and ${\mathcal F}$
can be written
\begin{eqnarray} \label{cross_cor}
\langle {\mathcal S}^2_{{\mathcal F} \mu}
\rangle\overline{\mathcal F}
= \frac{1}{N_{sn}}\sum_{z_s} \frac{1}{2} \int^{z_s}_0 dz~w_1(z,z_s) \frac{(1+z)^3W_l(z)}{d_L(z)^2} \int^\infty_{0}dL~L\phi(L,z)\Phi(L,z) 
\int_0^\infty dy~y\, w_2(y) 
\\ \nonumber \times \int^\infty_{-\infty} dr~
\xi_{\Phi\delta}(\sqrt{y^2+r^2},L,z).
\end{eqnarray}
where the last radial integral is over proper distance not comoving
distance.  
The luminosity function -- mass correlation function is defined by
\begin{eqnarray}
\xi_{\Phi\delta}(|\vec{x}_1-\vec{x}_2|,L,z)\equiv \langle
\delta\Phi(\vec{x}_1,L;\delta_{\vec{y}}) \delta(\vec{x}_2) \rangle
\end{eqnarray}
where all quantities are evaluated at the epoch corresponding to the 
redshift $z$.

\section{Modeling the Signal}
\label{sec:Model-Sign-Noise}
For estimates of the lensing signal, structure in the DM
distribution will be modeled as consisting of two components.  First
each galaxy has a DM halo surrounding it whose properties are correlated
with observable properties of the galaxy.  In addition, the galactic
halos are clustered into clusters which contain additional mass.
The lensing contributions from each of these components are considered 
separately in the next two subsections.

\subsection{Galaxy Halos}
\label{sec:Galaxy-Halos}
Let us make
the simplifying assumption that each galaxy sits at the
center of a massive halo so that $\delta\Phi(\vec{x},L) = -1$ everywhere
but in these locations.  In addition, it will be assumed that all the
positions and properties of the halos are uncorrelated with each other,
but directly related to the observable properties of the tenant galaxy.  
These assumptions are probably incorrect in detail, but they will serve well as a
statistical model whose validity can be tested by the observations.  The
additional mass contained in groups and clusters of galaxies will be
handled separately.

In this galaxy halo model the correlation function is given by
\begin{eqnarray}
\xi_{\Phi\delta}(r,L,z)=\delta_h(r,L,z)=\frac{\rho_h(r,L,z)}{\overline{\rho}_o(1+z)^3}-1
~,\\
 \int^\infty_{-\infty} dr~\xi_{\Phi\delta}(\sqrt{y^2+r^2},L,z)
=\xi_{g\Sigma}(y,L,z) =
\frac{\langle \Sigma_h(y,L,z) \rangle_\Theta}{\overline{\rho}_o(1+z)^3}
\label{xi_halo} \end{eqnarray}
where $\rho_h(r,L,z)$ is the density of the halo which surrounds a galaxy
of luminosity $L$ at redshift $z$ and $\Sigma_h(y,L,z)$ is its surface
density.  The average is over the orientation and substructure of the
halo.

For the purposes of estimating the signal the surface density of a
galactic halo will be modeled by
\begin{eqnarray}
\Sigma(y)=\frac{V_c^2}{2G} \frac{1}{y(1+y/r_c)^\gamma} ~~;~~~~
V_c = V_* \left( \frac{L}{L_*} \right)^\beta  ~~,~~ r_c = r_* \left
( \frac{L}{L_*} \right)^\alpha.
\label{halo}
\end{eqnarray}
This profile matches a singular isothermal sphere at small $y$ and then
drops away more rapidly beyond $r_c$ with an adjustable slope.  The
central velocity dispersion is related to the luminosity by the
Tully-Fisher relation, $\beta\simeq 1/4$.  The scale size, $r_c$, is
related to the luminosity by a less certain relation.  In what follows
$\alpha=1/2$ which in the case of $\beta= 1/4$ gives a constant
mass to light ratio for the isothermal part of the halos.  In addition 
this makes $r_c$ proportional to the radius of the visible
galaxy in an idealized case where the surface brightness is a constant.
The luminosity function is taken to be a Schechter function $\Phi(x
L_*)= \Phi_* x^{-1.07} e^{-x}$ with $L_*=10^{10} h^{-2} L_\odot$ and
independent of redshift.  The weighting function will be 
\begin{eqnarray}
w_1(z,z_s)=W_l(z,z_s) d_L(z)^2 ~~~,~~~ 
w_2(y)=\left\{ \begin{array}{cl}
\frac{1}{\pi (R^2-R_{min}^2)} & R_{min}<y<R \\
0 & y>R~,~y<R_{min}
\end{array} \right.
\label{window}
\end{eqnarray}
This is not necessarily the optimal weighting, but it does seem to give
high signal to noise without prejudicing the result by assuming a
specific halo profile.  With this weighting function ${\mathcal
S}^2_{{\mathcal F}\mu}(R)$ is directly related to the average surface
density in the annulus $R_{min}<y<R$ surrounding a galaxy.

In the above the weak lensing approximation has been used.  More
generally ${\mathcal S}^2_{{\mathcal F}\mu}(R)$ is related to the
average magnification within the annulus.  However the
probability of forming an image is not uniformly distributed on the
image plane due to magnification bias.  The result of this is that
${\mathcal S}^2_{{\mathcal F}\mu}(R)\propto \langle \mu \rangle_{sn}-1=
\langle \mu^{-1} \rangle_x^{-1}-1$ where $\langle\dots\rangle_x$ is the
average over the annulus $R_{min}<y<R$.  Weak lensing will be assumed
throughout this paper so this distinction in the interpretation of
${\mathcal S}^2_{{\mathcal F}\mu}(R)$ will not be important.

\subsection{Extragalactic Clustering}
\label{cluster_component}

It is well established that in the central regions of galaxy clusters
most of the mass is not associated with individual galaxies but resides
in intergalactic space.  This is confirmed by observations of gravitational lensing using galaxy
shear \markcite{1998ApJ...499..600}({Natarajan} {et~al.} 1998), the Sunyaev-Zel'dovich effect and
X-ray emission.  This extra cluster mass can have significant lensing
effects and conversely any information that can be gained through
lensing on the structure of clusters would contribute to our
understanding of cluster formation and evolution.

To model the effects of extragalactic mass - mass outside of any
galactic halo -  I treat the galaxy halos as
being embedded in larger halos with a universal profile but of varying size
and mass.  Only the component of extragalactic mass that resides in
virialized halos will be considered because any other component will not
contribute appreciably to the lensing considered here.
Because $\delta\mu \propto \Sigma$ to first order the
contribution of the extragalactic mass to lensing can then 
be calculated separately and simply added to the contribution from galactic
halos.  This will not be true for strongly lensed SNe, but these are
comparatively rare and will not be considered here.  I will call the extragalactic contribution the cluster
contribution although it will include extragalactic halos with masses well
below what is normally called a galaxy cluster.  The division between
galactic and cluster mass is somewhat arbitrary in any particular case,
but it does have a meaning in a statistical sense.   By comparing
cluster galaxies with the `average' 
galaxy the extra mass can be attributed to the cluster or group.
This separation is more direct and less model dependent than it is for
shear measurements \markcite{priya99,1998ApJ...499..600}(see Natarajan, Kneib, \& Smail 1999; {Natarajan} {et~al.} 1998) owing to
the magnification being directly proportional to the surface density.

The relative mass density of an extragalactic halo is written
$\delta_c(\vec{x},M)$ and its projection onto two dimensions is
$\delta\Sigma_c(\vec{y},M)$.  If the density of galaxies is assumed to
be proportional to the overdensity the correlation on the sky between
galaxies and surface density in halos with masses between $M$ and
$M+dM$ is given by
\begin{eqnarray}
\xi^{cl}_{g\Sigma}(y,M) &=& f_\Sigma(M) f_g(M) \int \frac{d\theta_y}{2\pi} \int d^2 x~
\delta\Sigma_c(\vec{x},M)\delta\Sigma_c(\vec{x}-\vec{y},M) / \int d^3 x~
\delta_c(\vec{x},M) \label{xi_c_1} \\
&=& \frac{f_\Sigma(M) f_g(M)}{\tilde{\delta\Sigma}_c(0,M)} \int \frac{dk k}{2\pi}
J_o(k y) ~| \tilde{\delta\Sigma}_c(k,M) |^2 \label{xi_c_2} 
\end{eqnarray}
where $f_g(M)$ is the fraction of galaxies in halos within this mass
range and $f_\Sigma(M)$ is the fraction of the cluster surface density that
is not in any galaxy halo.  The first integral in (\ref{xi_c_1}) is an average over the direction of
$\vec{y}$ or equivalently the orientation of the halo.
It is often more convenient to calculate the correlation function from
the Fourier transform of the halo profile,
$\tilde{\delta\Sigma}_c(k,M)$.  If the weight function $w_2(y)$ is 
given by~(\ref{window}) with $R_{min}= 0$, a top hat,
the relevant quantity is the average correlation function within a
circle of radius $R$,
\begin{eqnarray}
\overline{\xi^{cl}_{g\Sigma}}(M,R)=\frac{f_\Sigma(M) f_g(M)}{\pi R
\tilde{\delta\Sigma}_c(0,M)}\int dk~J_1(k R)~|
\tilde{\delta\Sigma}_c(k,M) |^2.
\end{eqnarray}
The $J$'s are Bessel functions.

Computationally the problem can be simplified when the halo profile
obeys the scaling law
\begin{eqnarray}
\tilde{\delta\Sigma}_c(k,M)=\delta_o(M)
r_{cl}(M)^3\tilde{\delta\Sigma}_c\left(r_{cl}(M)k\right)~~;~~\xi^{cl}_{g\Sigma}(y,M)=f_\Sigma(M) f_g(M)\delta_o(M) r_{cl}(M)~ \hat{\xi}^{cl}_{g\Sigma}\left(\frac{y}{r_{cl}(M)}\right)
\end{eqnarray}
where $r_{cl}(M)$ is the scale length of the halo and $\delta_o(M)$ is a
density scale.  Many popular profiles are of this generic type.  The scaling
implies a contribution to ${\mathcal S}^2_{\mathcal F \mu}$ from
extragalactic halos of 
\begin{eqnarray}\label{S_cluster}
\langle {\mathcal S}^2_{{\mathcal F} \mu}\rangle
= \frac{1}{\overline{\mathcal F}N_{sn}}\sum_{z_s} \frac{1}{4\pi} \int^{z_s}_0 dz~w_1(z,z_s) \frac{(1+z)^3W_l(z)}{d_L(z)^2} \int^\infty_{0}dL~L\phi(L,z)\Phi^{cl}(L,z) \\ \nonumber \times
\int^\infty_{M_{min}} dM~ f_\Sigma(M,z) f_g(M,z)\delta_o(M,z)
r_{cl}(M,z)~
\overline{\hat{\xi}^{cl}_{g\Sigma}}\left(\frac{R}{r_{cl}(M,z)}\right).
\end{eqnarray}
This scaling allows one to calculate
$\overline{\hat{\xi}^{cl}_{g\Sigma}}(x)$ once and then integrate over
halo mass and redshift.  The mass $M_{min}$ is the mass of the smallest
halo that contains multiple galaxies.  The luminosity function is given a
superscript because there is a known segregation of galaxy types between
clusters and the field.  However in the calculations to follow I take
the cluster galaxies to have the same luminosity function and halo
distribution as the galaxies in the field.
\begin{figure}
\centering\epsfig{figure=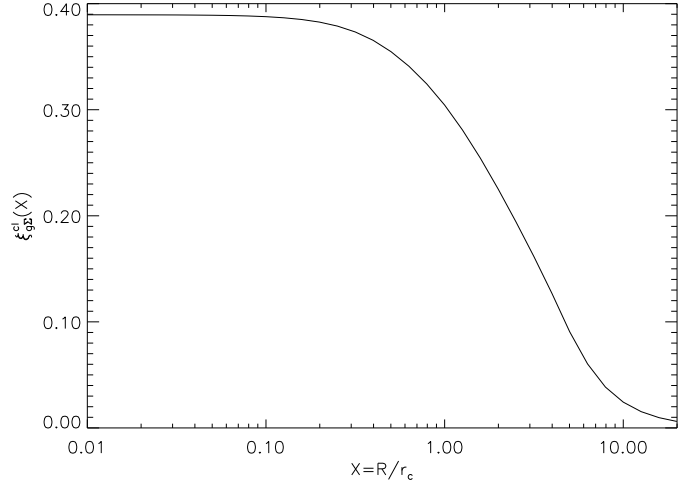,height=2.75in}
\caption[qiw]{The galaxy-surface density correlation function averaged over
a disk, $\overline{\hat{\xi}^{cl}_{g\Sigma}}(x)$.  The extragalactic halo is
taken to be of the NFW form (equation \ref{nfwhalo}).}
\label{xi_plot}
\end{figure}

For signal estimates I will assume the extragalactic clusters are of
the Navarro, Frenk and White (NFW) form \markcite{1997ApJ...490..493N,1996ApJ...462..563N}({Navarro}, {Frenk}, \&  {White} 1997, 1996):
\begin{eqnarray}
\delta_{cl}(r)=\frac{\delta_o}{(r/r_{cl}) \left(1+r/r_{cl}\right)^2}.
\label{nfwhalo}
\end{eqnarray}
The average galaxy-surface density correlation function,
$\overline{\hat{\xi}^{cl}_{g\Sigma}}(x)$, in this model is plotted in
figure~\ref{xi_plot}.  This function is effectively constant for
$x\simlt 0.1$ and zero for $x\simgt 10$.  

To calculate (\ref{S_cluster})
some assumptions concerning the population of halos and their galaxy content
must be made.  I will consider flat CDM models with scale free ($n=1$)
initial perturbations.
The parameters
$\delta_o(M,z)$ and $r_c(M,z)$ are related by $r_c=r_{200}(M)/c$ and
$\delta_o=\frac{200}{3\Omega_m(z)}c^3/\left[\ln(1+c)-c/(1+c)\right]$ where
$r_{200}(M)$ is the radius within which the mean density is 200 times
the background.  To simplify the procedure of calculating these
quantities the simple formula $c\simeq c_*(M/M_*(z))^{-0.2}$ is used.
The nonlinear mass scale is defined by the condition
$\Delta_o(M_*,z)=\delta_c(z)$ where the critical density is
$\delta_c=1.6865 \,\Omega_m(z)^{0.0055}$ for flat models.
These scalings reasonably reproduce the results of
\markcite{1996ApJ...462..563N}{Navarro} {et~al.} (1996).  
The normalizations for the $\Lambda$CDM model considered here
($\Omega_m=0.3$, $\Omega_\Lambda=0.7$) are $\sigma_8 = 1.14$ and
$c_*=7.9$.  For standard CDM ($\Omega_m=1$), $\sigma_8 = 0.6$ and
$c_*=10$.  The fluctuations on $8 h^{-1}\mpc$ scales, $\sigma_8$, is
consistent with observed galaxy abundances in both models \markcite{VL96}({Viana} \& {Liddle} 1996).


In populating the clusters with galaxies it is assumed that the fraction
of galaxies in clusters within a given mass range is equal to the
fraction of total mass which is contained in those clusters.  The mass
fraction can be calculated using the Press-Schechter prediction for the
mass function of halos.  This results in  
\begin{eqnarray}
f_g(M,z) = -\sqrt{\frac{2}{\pi}}\frac{\delta_c}{D(z)\Delta_o^2(M)}\frac{d\Delta_o}{dM}(M) ~\exp\left(\frac{-\delta_c^2}{2D(z)^2\Delta_o^2(M)}\right)
\end{eqnarray}
The galaxy halos are statistically the same in and out of
clusters.  The fraction of the mass in the cluster but not in a galactic
halo, $f_\Sigma(M,z)$, will not depend on $M$ in this case.  The 
assumption that the galaxy content of a cluster is proportional to mass
results in $f_\Sigma(z)=1-\rho_g(z)/\overline{\rho}(z)$ where
$\rho_g(z)$ is the total mass content of galaxies and their halos.  With
the static halo model considered here $f_\Sigma$ is independent of
redshift.  Typically
$f_\Sigma\sim 0.4$.  This number would increase if cluster galaxies are tidally
truncated as they undoubtedly are in reality.
\begin{figure}
\centering\epsfig{figure=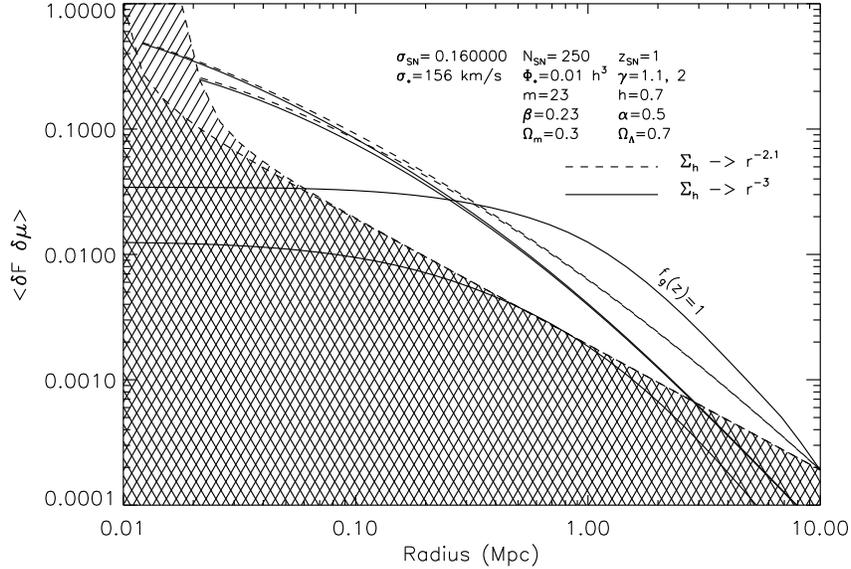,height=3.25in}
\caption[S_plot]{The correlation of light and magnification ${\mathcal
S}^2_{{\mathcal F}\mu}(R)$ for SNe at $z=1$.  The contributions from
galaxy halos and extragalactic halos are plotted separately.  The total
${\mathcal S}^2_{{\mathcal F}\mu}$ is the sum of the two components.
The galactic halo curves are the steeper ones starting in the upper left
corner.  Two different models which
differ in the logarithmic slope of the surface density at large
galactic radii are shown.  For each model two inner cutoffs are considered --
$R_{min}=10, 20\kpc$.  All SNe within $R_{min}$ of any galaxy are
excluded.  The two curves that flatten out on the left are the
contributions from clusters in the CDM model.  The one marked $f_g(z)=1$
represents the case where only SNe behind 
$M>10^{14}\msun$ clusters are selected.  The crisscrossed region represents the
uncorrelated noise contributed by uncertainties in SN peak luminosities
($\sigma_{sn}=0.16\mag$) and shot noise in the galaxy counts
for the case of 250 SNe (see section~\ref{sec:noise}).  The signal to noise
ratio at $R=200\kpc$ is $\simeq 4\sqrt{N_{sn}/250}$.  The limiting
magnitude of foreground galaxies is $m=23$.  Additional parameters are
printed on the plot.}
\label{S_plot}
\end{figure}
\begin{figure}
\centering\epsfig{figure=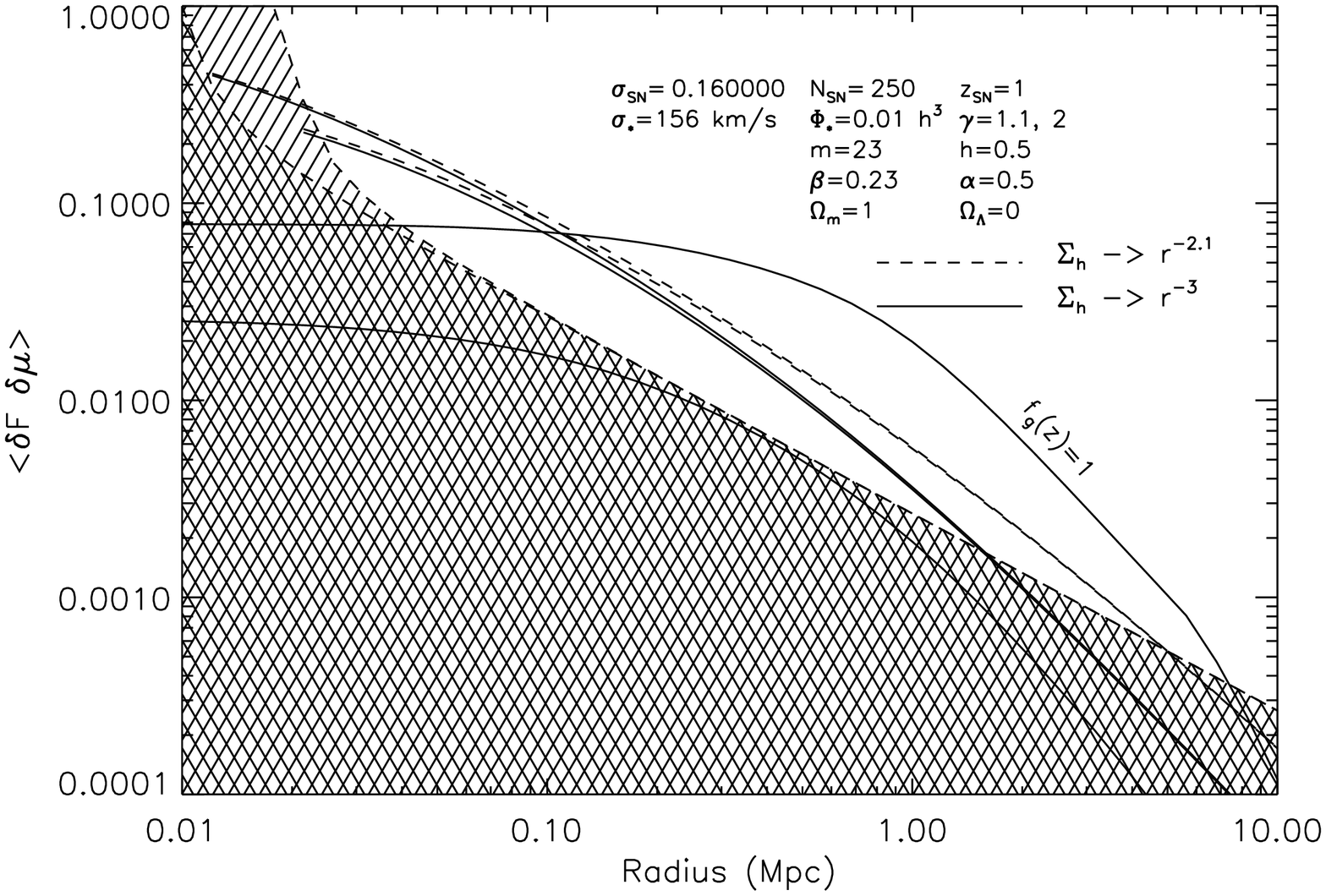,height=3.25in}
\caption[S_plot]{The same as in figure~\ref{S_plot} except $\Omega_m=1$,
$h=0.5$ and $\sigma_8=0.6$.}
\label{S_plot2}
\end{figure}

Figure~\ref{S_plot} shows the expected size of ${\mathcal S}^2_{\mathcal
F \mu}(R)$ for SNe at $z=1$ where $R$ is the radius of the weighting
function in~(\ref{window}). On top are curves representing the contribution of
galaxy halos.  In each case an artificial inner cutoff, $R_{min}$, is
made so that SNe whose line--of--sight comes within this distance of a
galaxy are excluded in order to avoid problems with obscuration.  This
is an adjustable parameter that is applied when analyzing the data.
${\mathcal S}^2_{\mathcal F \mu}(R)$ is then proportional to the average
surface density between $R_{min}$ and $R$.  Figure~\ref{S_plot} shows
that ${\mathcal S}^2_{\mathcal F \mu}(R)$ is relatively
independent of $R_{min}$ when $R\simgt R_{min}$.  The
inner part of these curves, beyond the $R_{min}$, depend only on the
isothermal part of the halo model and thus only on the
Tully-Fisher relation.  This could in fact be a way of determining if a
Tully-Fisher-Faber-Jackson type relation applies to halos.  Two different
halo models are shown with $\gamma=2$ and $1.1$ (halo model is given
in~(\ref{halo})).  These models start to diverge from each other at large $R$
($r_*=500\kpc$ in these models), but differentiating between them is
likely to require a large number of SNe both because of the small
difference and the contribution of cluster mass which starts to become
significant around the same radii.  It is clear in both models that
${\mathcal S}^2_{\mathcal F \mu}(R)$ deviates from isothermal as $R$
gets larger.  A measurement of the halo size scale, $r_*$, should be
possible.

The cluster contribution is represented in figure~\ref{S_plot} by the
lower curves which flattens out at small $R$.  The lower of these curves
is the contribution expected for a random sample of SNe.  In the present
model this contribution is fairly small relative to the noise.  This is
largely because of the small number of galaxies that reside in large
clusters.  For these calculations the minimum mass of a cluster is taken
to be $M_{min}=3 \times 10^{12}\msun$, but it is the larger clusters
which give most of the signal.  All else being fixed, the extragalactic
halo component will be proportional to $f_g(M)$ - the fractional galaxy
content of clusters.  In this way it is a measure of bias.  The noise
for a sample of 250 SNe is represented in figure~\ref{S_plot} by the
crisscrossed region.  With this number of SNe and the level of noise assumed
here the signal is $\sim 4.5\times\sigma$.  
a detection at the $4.5\,\sigma$ at $R\sim 0.1\mpc$.
The noise is discussed further in section~\ref{sec:noise}. 

The relative size of the cluster contribution can be
increased by selecting SNe -- or searching for SNe -- which occur
behind identified clusters.  The upper cluster curve in
figure~\ref{S_plot} marked $f_\Sigma=1$ shows the level of signal if all
the SN sight lines are within $\simeq 10 \mpc$ of the center of $M>
10^{14}\msun$ clusters.  Now the contribution from extragalactic material
is comparable or greater than the contribution from galaxy halos at
$R\sim 1\mpc$.  In actuality halos within a cluster should be tidally 
truncated which will make the halo curve fall more rapidly with $R$ and
the cluster curve will he higher due to the redistributed matter.  It
would be interesting to study the tidal striping process in this way.
It is strongly dependent on the formation history of clusters.
Of course the cluster mass threshold could be increased so that the cluster
lensing contribution is comparatively larger.  In this way the
substructure in galaxy clusters can be measured very directly.

Figure~\ref{S_plot2} shows the same ${\mathcal S}^2_{\mathcal F \mu}(R)$
plot with the background cosmology changed to $\Omega_m=1$.  The galaxy halos
are the same as in figure~\ref{S_plot}.  This galaxy halo signal is
almost independent of cosmology.  Perhaps unexpectedly $\Lambda$ does
not have a large effect even when the galaxy halo model is held fixed.  
This is a result of using correlations between light and magnification rather
than a statistic based purely on the magnification probability
distribution such as the optical depth.  The magnification probability 
is a stronger function of the angular size distance and the density of
halos while ${\mathcal S}^2_{\mathcal F \mu}$ is normalized to fit the
observed density of galaxies.  The cluster
component is a little bit bigger in the high $\Omega_m$ models when the
power spectrum is normalized to observed cluster abundances.

The modeling of extragalactic halos done here is somewhat crude.  There
are many ways in which it could fail to represent reality in detail.  
The density of galaxies within a halo is not necessarily proportional to the 
halo density and the total number of galaxies in a halo may not be
proportional to the halo mass.  It is already known that galaxies of
different morphologies, colors and luminosities are biased with respect
to each other.  This would seem to indicate that these assumptions are
not entirely correct.  In addition, large clusters tend to have a central
elliptical galaxy which has not been included here.  This would increase
the statistical weight of the cluster contribution.  More sophisticated
modeling is certainly possible when the data warrants it.

\section{Noise in $\langle{\mathcal S}^2_{{\mathcal F} \mu}\rangle$ }
\label{sec:noise}

There are several different sources of noise that will affect the
precision and accuracy of any $\langle{\mathcal S}^2_{{\mathcal F} \mu}(R)
\rangle$ measurement.  First there is noise coming from the
dispersion in SN luminosities and the random fluctuations in foreground
light.  This noise is uncorrelated with the lensing signal.  The lensing
structures themselves will also contribute to the
noise through sample variance.  The density within radius $R$ is
concentrated toward the center of the halo.  The more concentrated it is
the less good an estimate of the average from a finite sample is likely to
be.  This problem is compounded if the DM halos are not smooth but
clumped into substructures.  In addition errors in the redshifts of
foreground galaxies will add noise.  
There are also observational errors and
biases arising from the light curve correction to the SN peak
luminosity, the host galaxy subtraction and the detection efficiency of
the SN search.  For lack of detailed information
on future observations I will not attempt to address the last categories
of noise, but I will estimate the importance of the other sources of error.

The variance in the correlation can be written
\begin{eqnarray}\label{sig_S}
\langle \left[{\mathcal S}^2_{{\mathcal F} m} \right]^2 \rangle
-\langle{\mathcal S}^2_{{\mathcal F} m} \rangle^2 = \frac{1}{N_{sn}\overline{\mathcal F}^2}
\left[ \sigma_{\mathcal
F}^2 \sigma^2_b +\sigma_{\mathcal
F}^2 \left( \frac{\Delta\overline{b}}{\overline{b}}\right)^2 + {\mathcal P}^4_{{\mathcal F} \mu} \right]
\end{eqnarray}
where $\Delta\overline{b}(\Omega_m,\Omega_\Lambda)$ is the difference
between the true average brightness and the one derived using the
parameters; $\Omega_m,\Omega_\Lambda$, etc.  The variance in SN
brightnesses caused by intrinsic differences and observational errors is
$\sigma^2_b$.  It is assumed that the
error in $\overline{\mathcal F}$ is small enough that it is
unimportant.  The last term represents the sample variance produced by the
lensing itself.  All of these quantities will be redshift dependent so
in a real sample they will be different for each SN.

Errors in $\overline{b}(\Omega_m,\Omega_\Lambda)$ should always be a 
subdominant.  This is because the cosmology will be
constrained by all of the SNe observed, including those at lower 
redshift, so it is expected that $\Delta\overline{b}^2 \ll
\sigma^2_b$.  Of course there is always the possibility that one has
chosen the wrong parameters to describe the real universe.  With enough
SNe one could directly calculate $\overline{b}(\Omega_m,\Omega_\Lambda)$
within redshift bins.  This would make the result independent of
cosmology at the expense of increasing the
$\overline{b}(\Omega_m,\Omega_\Lambda)$ term.

\begin{figure}
\centering\epsfig{figure=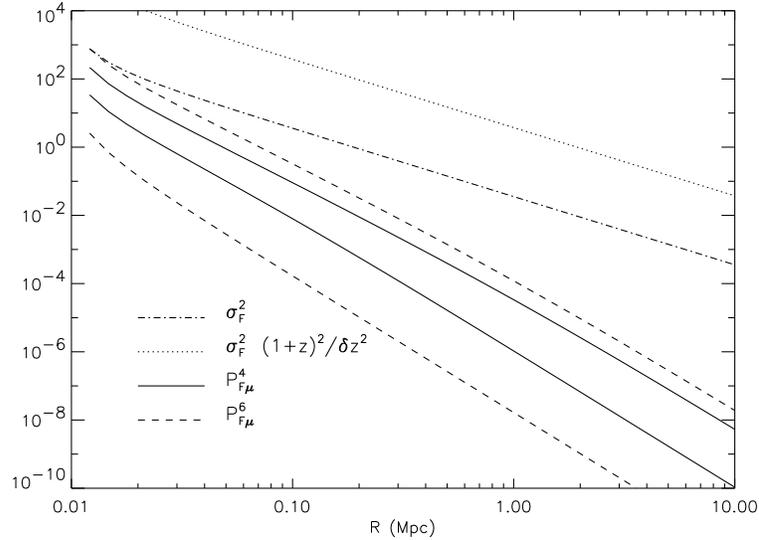,height=3in}
\caption[noise]{The major sources of noise for the statistics ${\mathcal
S}^2_{{\mathcal F}\mu}$ and ${\mathcal R}^2_{{\mathcal F}\mu}$.  These
are all calculated with the parameters used for figure~\ref{S_plot}.  
The dotted curve is the noise just from errors in redshift times
$(1+z)^2/\delta z^2$ which is regarded as a constant.  The dot-dashed
curve is the noise from shot noise in the galaxy counts.
For each pair of ${\mathcal P}^n_{{\mathcal F}\mu}$ curves the lower one
is without substructure and the higher one has substructure.  The
substructure clumps, as described in the text, are truncated isothermal
spheres with radius $R_{sub}=0.1\kpc$ and velocity dispersion
$v_c=15\kms$.  The magnification cutoff for ${\mathcal P}^4_{{\mathcal
F}\mu}$ and ${\mathcal P}^4_{{\mathcal F}\mu}$ is set at
$\delta\mu=10$.  All quantities are dimensionless.
}
\label{noise}
\end{figure}

The variance of ${\mathcal F}$ is best estimated directly from data.
However, to assess the noise levels expected in a hypothetical
experiment it is useful to make a semi-empirical model for the variance
$\sigma^2_{\mathcal F}$.  If it is assumed that the internal properties
of different galaxies are uncorrelated the variance is
\begin{eqnarray}
\sigma^2_{\mathcal F} & = & \langle {\mathcal F}^2 \rangle -
\overline{\mathcal F}^2 \nonumber\\
&=&\frac{2\pi}{H_o}\int_0^{z_s} dz~\frac{w_1(z,z_s)^2(1+z)^2}{\left[4\pi
d_L(z)^2\right]^2 E(z)} \left\{
\int^\infty_{0}dL~L^2\phi(L,z)\Phi(L,z) \int_0^{\infty} dy~y~w_2(y)^2
\right. \label{sigF2} \\ & & 
\begin{array}{r}
\left. + \frac{(1+z)^2}{2\pi} \int^\infty_{0}dL~L\phi(L,z)\Phi(L,z)
\int^\infty_{0}dL'~L'\phi(L',z)\Phi(L',z)
\int d^2y_1 \int d^2y_2 ~w_2(y_1)w_2(y_2)  \right. \\ \left. \times
\int^\infty_{-\infty}dr~ \xi_{gg}\left( \sqrt{\Delta y^2 + r^2}
,z\right) \right\}
\end{array}
\label{sigF1}\end{eqnarray}
The first term (\ref{sigF2}) is from shot noise in galaxy number
counts.  This expression can be derived by considering the variance in
the galaxy number count within a region of space small enough to contain
at most one galaxy.  The probability of a galaxy being
in this volume is given by equation~(\ref{galaxy_prob}).  Then a Poisson
distribution implies
\begin{eqnarray}
\left\langle \Phi(L)\Phi(L') \delta\Phi(\vec{x},L)
\delta\Phi(\vec{x}',L') \right\rangle =
\Phi(L)\delta(L-L')\delta^3(\vec{x}-\vec{x}')/(1+z)^3
\end{eqnarray}
where the positions $\vec{x}$ are measured in proper coordinates.
Term~(\ref{sigF2}) follows from this and is plotted in
figure~\ref{noise} for the low $\Omega_\Lambda$ flat model.

The second term in $\sigma^2_{\mathcal F}$, (\ref{sigF1}), is from the
clustering of galaxies.  It is known from observations that the
correlation function of galaxies is well approximated by
$\xi_{gg}(x)=\left(x/r_o\right)^{-\gamma}$ ($\gamma=1.77\pm0.004$ and
$r_o=5.4\pm1 h^{-1}\mpc$) and independent of redshift if $x$ is the
comoving distance.  With this level of clustering $\sigma^2_{\mathcal
F}$ is dominated by shot noise for relevant range of $R$ so clustering
contribution will be ignored.

In addition to shot noise the estimate of ${\mathcal F}$ is limited by
errors in the estimated redshifts of the foreground galaxies. 
These redshifts need to be estimated either photometrically or
spectroscopically.  This can cause errors in two ways.  If one is using
the weighting function~(\ref{window}) errors in 
redshift can cause $w_1(z_i,z_s)$ to be misestimated and it
can cause a 
galaxy that is outside the volume allowed by $w_2(y)$ to be
misidentified as inside that volume and vice versa.  If necessary the
latter problem can be avoided entirely by using an angular
cutoff in $w_2(y)$ instead of a proper distance cutoff as in~(\ref{window}). 
This is done at the expense of being able to interpret
$\langle {\mathcal S}^2_{{\mathcal F}\mu}(R) \rangle$ as a measure of the
average surface density within distance $R$ of a galaxy.  An estimate of the uncertainty in ${\mathcal F}$ caused by errors 
in $w_1(z,z_s)$ can be found by replacing
$w_1(z,z_s)$ in~(\ref{sigF2}) with $\frac{\delta z}{(1+z)} (1+z)
\frac{dw_1}{dz}(z,z_s)$ where $\delta z/(1+z)$ is considered constant.
This is also plotted in figure~\ref{noise}.  In the case considered here
$\sigma_{redshift} \simeq 10 \delta z/(1+z) \sigma_{shot\,noise}$.  At
this time photometric redshifts have an accuracy of $\delta z \sim
0.1-0.2$ for the relevant redshift range \markcite{astro-ph/0003380}(Bolzonella, Miralles, \&  Pello' 2000).  If photometric redshifts are used this noise
is comparable to the shot noise contribution.  With spectroscopic
redshifts this noise would be unimportant.  In figures~\ref{S_plot}
and~\ref{S_plot2} only the shot noise times the SN luminosity noise is shown.

The size of the sample variance depends on the structure of the halos.  If the
halos are very centrally concentrated, asymmetric and/or clumpy the
average ${\mathcal S}^2_{{\mathcal F}\mu}$ will strongly depend on a
small number of SNe which happen to be observed through high density
region.  In this case ${\mathcal S}^2_{{\mathcal F}\mu}$  may converge
to its mean value very slowly and a
prohibitively large number of SNe may be required to estimate it.  
Ignoring correlations between halos the sample variance can be written
\begin{eqnarray}\label{Pn}
{\mathcal P}^n_{{\mathcal F} \mu}= \frac{H_o^{n-3}}{\overline{\mathcal F}^2_{{\mathcal F} \mu}}
\int^{z_s}_0 dz~ w_1(z,z_s)^2 \frac{(1+z)^n E(z)^{n-3} W_l(z,z_s)^{n-2}
}{\left[ 4\pi d_L(z)^2\right]^2 \overline{\rho}(z)^{n-2}} \\ \nonumber
~~~~~~~~~~~~~~~~~~~~~~\times \int^\infty_{0}dL~L^2\phi(L,z)\Phi(L,z) 
\int d^2y~w_2(y)^2 B^{n-2}(L,z,z_s,y) 
\end{eqnarray}
with $n=4$ (the $n \neq 4$ will be useful later).

In the weak lensing limit
$B^{n}(L,z,z_s,y)=\langle\Sigma(L,y)^n\rangle$.  If the halo structure
is perfectly correlated with the galaxy luminosity $B^{n}(L,z,z_s,y)=
\overline{\Sigma}(L,y)^n$.  There is undoubtedly some scatter in the
luminosity--surface density relation which will increase ${\mathcal
P}^4_{{\mathcal F} \mu}$.  The scatter is not known, but if we take the
scatter in the Tully--Fisher relation as a guide ${\mathcal
P}^4_{{\mathcal F} \mu}$ will be about $10\%$ larger than the perfect
correlation value.  This should be added to the lower solid curve
in figure~\ref{noise}.  The sample variance would then be smaller than the
first term in~(\ref{sig_S}).  However the scatter could well be larger
than it is for the Tully--Fisher relation in which case one might hope
to measure it with higher order statistics.  This is discussed in
section~\ref{sec:Light-magn-magn}.

Another possibility is that the there is a significant amount of
substructure in the dark matter halos.  In this case I will resort to an 
approximation for $B^{n}(z,z_s,y)$ to simplify calculations.
In order to preserve the average angular size distance 
the average magnification at a distance $y$ must be the same with or without
substructure.  A model for substructure lensing should respect this
constraint and take into account the strong lensing that may be
important if subclumps have singular cores.  I consider the case where
all the mass in the halo is contained in randomly placed clumps each
with average surface density $\overline{\Sigma}_{sub}$.  The probability
of a clump being in the line--of--sight is then
$p=\overline{\Sigma}(y)/\overline{\Sigma}_{sub}$.  Outside the clumps
the magnification is zero (it depends on shear only at second order).  
When the halo density is large the possibility of two clumps eclipsing
the line--of--sight must be taken into account.  In the weak lensing limit
$B^{n}(z,z_s,y) \simeq p(1+2p)\,\langle \overline{\Sigma}_{sub}^n
\rangle$ where only the terms of highest order in
$\overline{\Sigma}_{sub}$ have been retained.
To take strong lensing into account I
renormalize this by $\langle \delta\mu^n \rangle/\langle \delta\mu \rangle^n$.
The result is
\begin{eqnarray}\label{Bn}
B^n(L,z,z_s,y)  = 
\left\{ 
\begin{array}{cl}
\overline{\Sigma}(y)^n &, \mbox{ no substructure} \\
\frac{\langle \delta\mu(z)^n \rangle}{\langle \delta\mu(z) \rangle^n}
\overline{\Sigma}_{sub}^{n-1} \overline{\Sigma}(L,y)
\left(1+2\overline{\Sigma}_{sub}^{-1} \overline{\Sigma}(L,y) \right)  &, \mbox{ substructure}
\end{array} \right.
\end{eqnarray}
If a magnification cutoff is imposed to make $\langle\delta\mu^n
\rangle$ converge, $\langle\delta\mu\rangle^n$ is still calculated
without this cutoff so that the proper normalization is retained.

To model the magnification within a subclump I use the solution for an
isothermal sphere cutoff at an outer radius, $R_{sub}$.  The
characteristic length scale for such a lens is $x_o=4\pi v_c^2
D_lD_{ls}/D_s$.  The magnification, including both images, is then a
function of only the rescaled impact parameter $x=y/x_o$:
\begin{eqnarray}
\delta\mu=\left\{ 
\begin{array}{cl}
(2-x)/x &,~x<1 \\
1/x &,~ 1<x<x_m \\
0 &,~ x>x_m
\end{array} \right.
\end{eqnarray}
The second moment is logarithmically divergent so a maximum
magnification cutoff must be imposed to calculate~(\ref{Pn}).  This
cutoff can be set high enough that the probability of there being a SN in the
observed sample with a magnification this high is very small.
The divergence of the moments is of course not real.  In actuality the
finite size of the source puts a limit on the magnification.

The size of dark subclumps is not very well constrained by observations,
but they do appear in simulations.  \markcite{1999ApJ...524L..19M}{Moore} {et~al.} (1999a) find a
large number of subclumps in their CDM simulations - many more then are
accounted for by visible dwarf galaxies.  For these simple estimates I
take the velocity dispersion to be $v_c=15\kms$ and the size to be
$100\mbox{ pc}$.
This is an extreme model since all the mass is in rather large
subclumps, but it has the advantage of simplicity and gives a
conservative estimate for the noise.  The sample variance,
 ${\mathcal P}^4_{{\mathcal F}\mu}(R)$, with and without this
substructure is plotted in figure~\ref{noise}.  The cutoff is set at
$\delta\mu>10$.  In this model there is about a $4\times10^{-4}$ chance
of a SN at $z=1$ being magnified above this limit.

With the variance expected in the corrected SN luminosities the
uncorrelated noise, the first term in~(\ref{sig_S}), generally dominates
over the sample variance.  At small $R$ the sample variance may be
comparable to the uncorrelated noise, but to some extent this is a result
of the model used here.  One would expect the subclumps to break up when the
density is high which is not reflected in the model.  
Another reason to think that this is an overestimate of the sample
variance is that the density profiles of real CDM clumps are probably
not as steep as $\rho\propto r^{-2}$ at small $r$.  It has recently been
claimed that $\rho\propto r^{-1.5}$ is a better
approximation \markcite{1999MNRAS.310.1147M}({Moore} {et~al.} 1999b) and the NFW model has
$\rho\propto r^{-1}$.  In these cases  ${\mathcal P}^4_{{\mathcal
F}\mu}$ would be smaller.

There is one other outstanding problem that could still cause trouble for
this measurement.  It has been suggested that an exceptional kind of
dust could account for the observed faintness of high redshift SNe.  The dust
grains would need to be large enough to avoid the constraints on reddening
\markcite{1999ApJ...525..583A,1999ApJ...517..565P}({Aguirre} 1999; {Perlmutter} {et~al.} 1999) and quite uniformly
distributed in the intergalactic medium \markcite{astro-ph/0002422}(Croft {et~al.} 2000).  This
is not consistent 
with the properties of known dust, but it does remain a possibility.
The distribution of this dust would presumably be correlated with
galaxies so it would show up as an anti-correlation between foreground
light and SN brightness.  Looking for correlations might actually be the
best way of testing this hypothesis as well.

In summary it seems promising that $\langle {\mathcal S}^2_{{\mathcal
F}\mu}\rangle$ could be measured with $\sim 250$ at $z\simeq 1$.  If the
SNe are observed with a future satellite it is likely that the noise
will be smaller than is assumed here because in this case $\sigma_{sn}$ will be
smaller at high redshift.

\section{Higher Order Correlations}
\label{sec:High-Order-Corr}

The above discussion of sample variance leads to the question of whether 
dark subclumps in galaxy halos might be detectable through the gravitational 
lensing of SNe.  Such a detection would have important implications for
structure formation theories.  
Recent high resolution N-body simulations have revealed that the CDM
models have several apparent problems when compared to observations.
One such problem is that galaxy halos in simulations have much more
substructure or smaller clumps within them than observed galaxies have
observable dwarf satellites \markcite{1999ApJ...524L..19M}({Moore} {et~al.} 1999a).  These
subclumps form 
early and fall into the larger halos without being destroyed by tidal
stripping and collisions.  Subclumps of dwarf galaxy size seem to be
over produced by about a factor of 50 in halos that should host
Milky Way--like galaxies.  Considering the successes of the CDM model
it seems reasonable to entertain the possibility that these subclumps
exist, but are dark either because they never formed stars or because
their star formation was shutoff early.\footnote{It is interesting to
note that if this is the case the present microlensing results do not
put very strong constrains on the fraction of the halo in MACHOs
\markcite{1996ApJ...464..218M}(see Metcalf \& Silk 1996). }  

In this section I investigate the two third-order correlation functions
of $\delta\mu$ and ${\mathcal F}$ and find that one is sensitive to this
substructure.  The other contains additional information on the
clustering of dark matter on extragalactic scales.

\subsection{Light-magnification-magnification Correlation}
\label{sec:Light-magn-magn}

In direct analogy to ${\mathcal S}^2_{{\mathcal F}\mu}$ I define the
statistic ${\mathcal R}^3_{{\mathcal F}\mu}$:
\begin{eqnarray}
{\mathcal R}^3_{{\mathcal F}\mu} \equiv  \frac{1}{N_{sn}} \sum_j
\frac{\Delta {\mathcal F}_j \Delta b_j^2 }{ \overline{\mathcal
F}_j~\overline{b}_j^2 }
 ~~~,~~~ \left\langle {\mathcal R}^3_{{\mathcal F}\mu}\right\rangle = \left\langle\delta{\mathcal F} \delta\mu^2 \right\rangle
\end{eqnarray}
Like ${\mathcal S}^2_{{\mathcal F}\mu}$ the average of this statistic
will be zero in the absence of lensing.

Assuming weak lensing the average value of ${\mathcal R}^3_{{\mathcal
F}\mu}$ is
\begin{eqnarray}
\langle {\mathcal R}^3_{{\mathcal F}\mu} \rangle \overline{\mathcal F}
= \frac{H_o}{2N_{sn}}\sum_{z_s} \int^{z_s}_0 dz~ w_1(z,z_s)
\frac{(1+z)^4 E(z) W_l(z)^2}{d_L(z)^2}\int^\infty_{0}dL~L\phi(L,z)\Phi(L,z) 
\int_0^\infty dy~y\, w_2(y) 
\\ \nonumber \times \int^\infty_{-\infty} \int^\infty_{-\infty} dr_1 dr_2~
\zeta_{\Phi\delta\delta}\left(y,\sqrt{y^2+r_1^2},\sqrt{y^2+r_2^2},L,z\right).
\label{Raverage}
\end{eqnarray}
\begin{eqnarray}
\zeta_{\Phi\delta\delta}(x_1,x_2,x_3,L,z) = \langle
\delta\Phi(\vec{x}_1,L;\delta_{\vec{y}})\delta(\vec{x}_2)
\delta(\vec{x}_3) \rangle \nonumber
\end{eqnarray}
${\mathcal R}^3_{{\mathcal F}\mu}$ is related to how fluctuations in
mass density, instead of just the mass density, are correlated with light.
Taking each galaxy halo to be independent and thin results in the third
order analog of equation~(\ref{xi_halo})
\begin{eqnarray}
\int^\infty_{-\infty} \int^\infty_{-\infty} dr_1 dr_2~
\zeta_{\Phi\delta\delta}\left(y,\sqrt{y^2+r_1^2},\sqrt{y^2+r_2^2},L,z\right) =
\frac{ \langle \Sigma_h(y,L,z)^2
\rangle_\Theta}{\overline{\rho}_o^2(1+z)^6}.
\label{zeta}
\end{eqnarray}

\begin{figure}
\centering\epsfig{figure=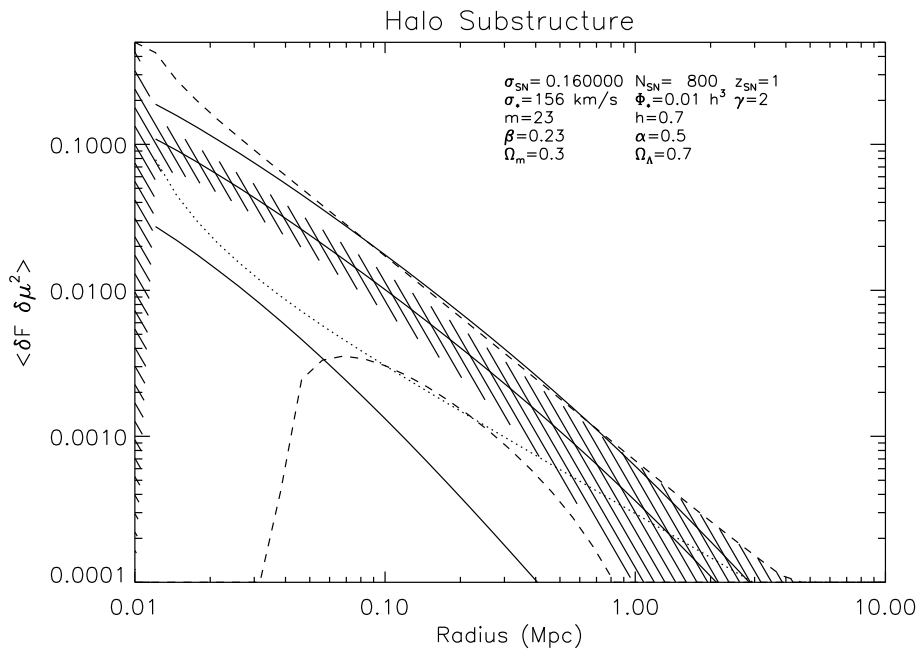,height=3.5in}
\caption[R_plot]{The three solid lines are the second order correlation
function ${\mathcal R}^3_{{\mathcal F} m}$.  The lowest of these is the
result with no halo substructure and perfect correlation of the halo with
the luminosity of its resident galaxy.  The dotted curve is the noise in
this case.  The scatter would need to increase ${\mathcal
R}^3_{{\mathcal F} m}$ by a factor of several for it to be measurable.
The other curves include substructure in the halos.
The top solid curve is the result with $\delta\mu > 10$ excluded.    All
the other curves are for $\delta\mu > 4$ excluded.
The inner crossed region represents the noise without sample
variance.  The dashed lines are the one-$\sigma$ noise with sample 
variances.  These are all calculated for 800 SNe at $z=1$.
}\label{R_plot}
\end{figure}

The second moment of the surface density will differ from simply the square
of the average halo profile because of scatter in the $L$--$\Sigma$ relation,
asymmetries and substructure.  Asymmetries may increase the moment by a
small factor.  Scatter $L$--$\Sigma$ relation could potentially be
larger -- perhaps a factor of a few.  Substructure could potentially
have a largest effect.   I use the substructure model discussed in
section~\ref{sec:noise} to estimate 
how sensitive ${\mathcal R}^3_{{\mathcal F}\mu}$ is to substructure.
The resulting calculation is plotted in figure~\ref{R_plot}.

The variance in ${\mathcal R}^3_{{\mathcal F}\mu}$ can be written
\begin{eqnarray}\label{Rvariance}
\langle \left[{\mathcal R}^3_{{\mathcal F} m} \right]^2 \rangle -\langle
{\mathcal R}^3_{{\mathcal F} m} \rangle^2 = \frac{1}{N_{sn}}
\left(\sigma_{\mathcal F}^2
\mu^4_\mu + {\mathcal S}^2_{{\mathcal F} \mu} {\mathcal
Q}^4_{{\mathcal F} \mu} + {\mathcal P}^3_{{\mathcal F} \mu} \mu^3_\mu +
{\mathcal P}^4_{{\mathcal F} \mu} \sigma^2_b + {\mathcal P}^6_{{\mathcal F} \mu} \right)
\end{eqnarray}
where ${\mathcal P}^n_{{\mathcal F} \mu}$ is defined in~(\ref{Pn}) and 
\begin{eqnarray}\label{Q}
{\mathcal Q}^4_{{\mathcal F} \mu}=\frac{H_o^2}{4\pi\overline{\mathcal F}}
\int_0^{z_s}dz~w_1(z,z_s) \frac{(1+z)^5 E(z)^2}{d_L(z)}W_l(z,z_s)^3
\int^\infty_{0}dL~L\phi(L,z)\Phi(L) \int d^2y~w_2(y) B^2(L,z,z_s,y).
\end{eqnarray}
The approximated value of $B^2(L,z,z_s,y)$ is given in~(\ref{Bn}).  The
moments of the distribution of intrinsic SN luminosities are 
$\mu^n_\mu$.  This distribution will be approximated as Gaussian in
which case $\mu^3_\mu=0$ and $\mu^4_\mu=3\sigma^4_\mu$.
In reality each of these factors will be different for each SN because
of differences in redshifts and observations so an average over SNe
will be done.

Unlike in the case of ${\mathcal S}^2_{{\mathcal F}\mu}$ the noise in
${\mathcal R}^3_{{\mathcal F}\mu}$ is usually dominated by sample
variance rather than SN luminosity errors and galaxy count noise.  As a
result both the signal and noise are strongly dependent on the structure
of the subclumps.  This is a result of the strongly peaked central
density of the subclumps.  Specifically the ${\mathcal P}^6_{{\mathcal
F} \mu}$ term in~(\ref{Rvariance}) tends to dominate.  If the cutoff in
$\delta\mu$ is high the sample variance will swamp the signal.  As can
be seen in figure~\ref{R_plot} if there is no substructure $\langle {\mathcal
R}^3_{{\mathcal F}\mu}\rangle$ is too small to be measured with 800 SNe
at $z\simeq 1$ because of sample variance.  Substructure increases the
signal, but also the sample variance.  A magnification cutoff can be
used to reduce the sample variance without reducing the signal as
significantly.  The cutoff is a censoring of the data in order to improve
the convergence of the remaining subsample.  The high magnification
events are rare so if the cutoff is at high magnification the subsample
is likely to contain all the data.  The subsample can be viewed as
sampling the area on the surface of the halo where the magnification is
below $\delta\mu_{max}$.  By adjusting $\delta\mu_{max}$ we can choose
to have a more precise result relating to a restricted area or a less
precise result relating to more of the area.

The sample variance may not be as big a problem in reality as it is in
this model of substructure.  Pure CDM simulations indicate that
the density in the clump cores is not as steep a function of radius as
in the singular isothermal sphere model
\markcite{1996ApJ...462..563N,1999MNRAS.310.1147M}({Navarro} {et~al.} 1996; {Moore} {et~al.} 1999b).  The density is more
likely to go as $r^{-1.5}$ or $r^{-1}$ rather than $r^{-2}$.  This
reduces $\langle {\mathcal R}^3_{{\mathcal F}\mu} \rangle$, but it
reduces the sample variance more significantly.  On the
other side, galaxy halos are not likely to be made entirely of such
large subclumps.
\markcite{1999ApJ...524L..19M}{Moore} {et~al.} (1999a) find that subclumps in their simulations have
a mass distribution $dn/dm\propto m^{-2}$ down to their
resolution limit.  Smaller or fewer subclumps would reduce the signal by 
approximately the covering fraction of the clumps.

\subsection{Light-light-magnification correlation}
\label{sec:Light-light-magn}

Finally, for completeness, there is the light-light-magnification
correlation function
\begin{eqnarray}
{\mathcal I}^3_{{\mathcal F}\mu}\equiv \frac{1}{N_{sn}} \sum_j 
\frac{\Delta{\mathcal F}_j^2 \Delta b_j  }{ \overline{{\mathcal F}}_j^2
~\overline{b}_j}
~~~~,~~~~
\left\langle {\mathcal I}^3_{{\mathcal F}\mu}\right\rangle = \left\langle
\delta{\mathcal F}^2 \delta\mu \right\rangle
\end{eqnarray}
\begin{eqnarray}\label{I3}
\begin{array}{c} \langle {\mathcal I}^3_{{\mathcal F}\mu} \rangle\overline{\mathcal F}^2
= \frac{1}{N_{sn}}\sum_{z_s} \int^{z_s}_0 dz~ w_1(z,z_s)^2 \frac{ (1+z)^6
W_l(z)}{\left[4\pi d_L(z)^2\right]^2}
\int^\infty_{0}dL~L\phi(L,z)\Phi(L,z)\int^\infty_{0}dL'~L'\phi(L',z')\Phi(L',z)
 \\ ~~~~~~~~~~~~~\times
\int^\infty_{-\infty} d^3r\int^\infty_{-\infty} d^3r' 
~w_2(r_\perp)w_2(r'_\perp)~\zeta_{\Phi\Phi\delta}\left(|\vec{r}|,|\vec{r'}|,|\vec{r}-\vec{r'}|,L,L',z\right)
\end{array}
\end{eqnarray}
\begin{eqnarray}
\zeta_{\Phi\Phi\delta}(x_1,x_2,x_3,L,L',z) = \langle
\delta\Phi(\vec{x}_1,L;\delta_{\vec{y}})\delta\Phi(\vec{x}_2,L';\delta_{\vec{y}})
\delta(\vec{x}_3) \rangle
\end{eqnarray}
The integrals over the third order correlation function
$\zeta_{\Phi\Phi\delta}(x_1,x_2,x_3,L,L',z)$ can be expressed as the sum
of two terms.  One is a shot noise term resulting from galaxy halos
taken individually.  Like $\langle {\mathcal S}^2_{{\mathcal
F}\mu}\rangle$ this term is proportional to the galaxy-surface-density
correlation function.  Then there is a term from correlations between
different galaxies which is proportional to the
galaxy-galaxy-surface-density correlation function.  This term falls off
less rapidly with $R$ so that at large radii $\langle {\mathcal I}^3_{{\mathcal
F}\mu}\rangle$ is related to the average surface density associated with
multiple galaxies.

It was shown in section~\ref{cluster_component} that by selecting
(or searching for) SNe that are seen through galaxy clusters ${\mathcal
S}^2_{{\mathcal F}\mu}(R)$ can be used as a probe of cluster matter.
The selection process would likely have a 
rather high cluster mass threshold.  To probe smaller mass clusters and
halos a more systematic approach might be useful.
The statistic ${\mathcal I}^3_{{\mathcal F}\mu}(R)$ provides a more direct,
although somewhat noisier, way of probing extragalactic halo structure.  
This will not be feasible until a considerable number SNe have been
observed so no detailed
calculations of $\langle {\mathcal I}^3_{{\mathcal F}\mu} \rangle$ will
be attempted here.

\section{Comparison with the lensing of galaxies and quasars}
\label{sec:Comp-with-lens}

The gravitational lensing of galaxies is now an established field of
observational astronomy.  It is appropriate that the lensing of SNe be
compared with that of galaxies to argue for their independence and
complementarity.  Weak lensing can be detected either through the
distortion of galaxy images or variations in their number counts.  The
former is a measure of the shear field and the latter is a measure of
the magnification.
Galaxy--galaxy lensing, the technique most similar to
the one discussed here, is where the image distortion of background
galaxies is correlated with their projected distances from foreground
galaxies.  This results in a measurement of the 
average shear as a function of distance from a galaxy.  Such
measurements have been done or attempted by several groups
\markcite{astro-ph/9912119,1998ApJ...503..531H, 1996ApJ...466..623B,
1996MNRAS.282.1159G, 1984ApJ...281L..59T}(Fischer {et~al.} 1999; {Hudson} {et~al.} 1998; {Brainerd} {et~al.} 1996; {Griffiths} {et~al.} 1996; {Tyson} {et~al.} 1984).
The tangential shear averaged around any circle on the sky
of radius $\theta$ is given by the remarkably simply relation
$\overline{\gamma}_t(\theta) =
\overline{\kappa}(<\theta)-\overline{\kappa}(\theta)$ where
$\kappa(\theta) = \Sigma(\theta)/\Sigma_c(z,z_s)$ is the convergence.  The average
convergence within the circle is $\overline{\kappa}(<\theta)$ and the
average on the circle is $\overline{\kappa}(\theta)$.  In the
weak lensing limit $\delta\mu(\vec{\theta}) = 2\kappa(\vec{\theta})$ so
for a singular isothermal sphere $\overline{\gamma}_t(\theta) =
\overline{\kappa}(\theta)$ and $\delta\mu(\vec{\theta}) \simeq 2
\overline{\gamma}_t(\theta)$.  If the mass profile falls off more
rapidly than isothermal, as it must for the mass to be finite,
$\delta\mu(\vec{\theta})$ will fall off more steeply than
$\overline{\gamma}_t(\theta)$.  For this reason measuring $\delta\mu(r)$
with SNe might be a better method for finding the size of galaxy
halos.  Galaxy-galaxy lensing has not been able to put any clear
constraint on the size of halos.  The transition between galaxy halos
and intercluster material might also be more distinct with SNe.

The shear and magnification measurements could also be combined;
galaxy--galaxy lensing constraining mostly the mass scale of halos and SN
lensing the size scale.  In this way degeneracies and systematic errors could be
significantly reduced.  The signal to noise per source is actually
better for SNe than for galaxies.
The rms ellipticity of galaxies is $\langle \epsilon^2 \rangle^{1/2}
\simeq 0.3$ so for isothermal lenses roughly 10-25 times the number of
background galaxies as SNe are needed to attain the same signal to
noise.  This factor brings the number of require SNe into the range that
is already being considered for measuring cosmological parameters.  

As discussed in section~\ref{sec:Light-magn-magn} there is the
possibility of measuring substructure in galactic halos with SNe.
Because the image sizes of the background galaxies are generally much larger
than the size of subclumps galaxy-galaxy lensing will not be able to
measure any realistic level of substructure.  In addition, the
contributions to the shear from the halo as a whole and the subclumps
are incoherent.  In contrast the lensing of SNe is sensitive to
substructure masses all the way down to a small fraction of a solar mass.

Another use for the lensing of galaxies is to actually map the shear
field.  Because each galaxy has an intrinsic ellipticity the shear can
only be measured in cells which contain multiple galaxies.  This limits
the angular resolution on which the shear field can be mapped.
The same is true for magnification
bias where the number density of background galaxies is related to the
magnification.  The lensing of SNe is unlikely to ever be useful for
this kind of study because the number density on the sky will always be
significantly lower than the number density of galaxies.  However, SNe
may be used to remove the mass sheet degeneracy that exists in shear
maps of galaxy clusters \markcite{kolatt98}({Kolatt} \& {Bartelmann} 1998).

Quasars have been used as sources in lensing studies for years.  They
are similar to SNe in that they are nearly point sources, but they are
not standard candles.  As a result lensing is only detectable
when multiple images are formed.  This is a rare occurrence and tells use
more about the central regions of halos than their outskirts.
Monitoring the light curves of multiple images does seem to be a
promising method for detecting compact objects.  Subclumps the size of 
dwarf galaxies however will have time scales that are to long to be
detected in this way.

\section{Discussion}

It has been shown that gravitational lensing can be measured by 
correlating SN brightnesses with the density of foreground galaxies.  
The result is a measure of the correlation of mass with light, the
bias, on $\sim \mpc$ scales.  The size and mass scales of galaxy halos 
should be measurable with a few hundred SNe at $z\sim1$.  The structure
of galaxy clusters can also be probed by selecting SNe that are viewed
through clusters.  Higher order correlations can be used to probe
substructure in galaxy halos.  In this case the convergence of the
result is strongly dependent on the form of the substructure that is 
present.  

The prospects for making these measurements seem very good.
The numbers of SNe and the noise levels discussed in this paper are
attainable, even conservative.  The proposed VISTA telescope is expected
to be able to find hundreds of SNe at redshifts $\simgt
1$.\footnote{home page: http://www-star.qmw.ac.uk/$\sim$jpe/vista/}  The
proposed SNAPSAT satellite could find thousands up to $z \simeq
1.7$.\footnote{home page: http://snap.lbl.gov/proposal/}   At high
redshift the major source of uncertainty in a SN luminosity comes from
subtracting the light of the host galaxy.  The superior resolution of a
satellite could reduce the variance in the corrected peak luminosities of
high redshift SNe to that of local SNe, $\Delta m \simeq 0.12\mag$ or better.
This would be a significant improvement on the $\Delta m \simeq 0.16\mag$ assumed
throughout this paper.  Gravitational lensing adds strongly to the motivations
for these projects.

\acknowledgments
I would like to thank D. Mortlock and P. Natarajan for very helpful
comments on this paper.



\end{document}